\documentstyle[12pt,doublespace,epsf]{article}

\newcommand{
\xtdots}{\hspace{3pt}\ddot{}\mbox{}\hspace{3.3pt}\dot{}\hspace{-6pt}x}
\newcommand{\xfodots}{
\hspace{3pt}\ddot{}\mbox{}\hspace{3.9pt}\ddot{}\hspace{-6pt}x}
\newcommand{\xfidots}{\hspace{3pt}\ddot{}\mbox{}\hspace{3.2pt}\dot{}
\mbox{}\hspace{3.5pt}\ddot{}\hspace{-7.5pt}x\hspace{2.4pt}}

\begin{document}
\thispagestyle{empty}

\mbox{}
\vspace{1cm}
\begin{center}
{\bf OPTIMUM PATHS FOR SYSTEMS SUBJECT TO INTERNAL NOISE } \\
\vspace{0.5cm}
{\it S. J. B. Einchcomb and A. J. McKane} \\
\vspace{0.5cm}
 Department of Theoretical Physics \\
University of Manchester \\ Manchester M13 9PL, UK \\
\end{center}

\begin{abstract}
We formulate the stochastic dynamics of a particle subject to internal
non-white (coloured) noise in terms of path-integrals. In the simplest case,
where the noise is exponentially correlated, the weak-noise limit is
characterised by optimum paths which are given by third order differential
equations. In contrast to systems subject to white noise or external
coloured noise, the overdamped limit for these systems is singular. We
analyse the origin of this behaviour. The whole formalism is generalised to
more general noise processes and the essential features are shown to be
similar to the exponentially correlated case.
\end{abstract}

\newpage
\pagenumbering{arabic}

\section{Introduction}

The phenomenological description of the statistical dynamics of a physical
system in terms of a set of Langevin equations is very familiar. However,
the frequently-made assumption that the noise in these equations is white
does not always hold. This happens when the motion of the particles of the
system depend on the history of the system, and not just on the instantaneous
values of position and velocity. Physical instances where this occurs include
chemical reactions
in solutions \cite{gah}, the migration of ligands in biomolecules \cite{han},
a mechanism for a glass transition \cite{leu}, the simulation of electrolytic
solutions \cite{ktb}, the atomic diffusion through a periodic lattice
\cite{tm,im} and others.

The derivation of the Langevin description for a system interacting with its
environment (a heat bath at a temperature $T$)
was put on a firm foundation by Zwanzig \cite{zwa}
and others. One assumes that the system plus environment is described by the
deterministic equations of classical mechanics (we shall not consider quantum
systems here). Integrating out the degrees of freedom associated with the
environment gives a set of equations for the system degrees of freedom which
contain the (random) initial conditions of the bath variables. In the simplest
case of a single particle of mass $m$ moving in a one-dimensional potential
$V(x)$, these equations have the form \cite{zwa}:
\begin{equation}
m\ddot{x}(t) + \int^t_{-\infty} dt' f(t - t') \dot{x}(t') + V'(x(t)) = \xi (t)
\label{langevin}
\end{equation}
where the noise $\xi$ is Gaussian with zero mean and correlator
\begin{equation}
\langle \xi (t) \xi (t') \rangle = k_B T f(t - t')
\label{noise}
\end{equation}
where $k_B$ is Boltzmann's constant. The Langevin equation (\ref{langevin})
is non-Markovian since the frictional term contains a memory function
$f(t)$. The strength of the dissipation will be denoted by $\alpha$, and
defined by
\begin{equation}
\alpha = \frac{1}{2}\int^{\infty}_{-\infty} dt f(t)
\label{diss}
\end{equation}
Under certain circumstances \cite{zwa} this memory function can be
well approximated by a delta-function and (\ref{langevin}) and (\ref{noise})
reduce to
\begin{equation}
m\ddot{x}(t) + \alpha \dot{x}(t) + V'(x(t)) = \eta (t)
\label{whlang}
\end{equation}
where the noise $\eta$ is now white:
\begin{equation}
\langle \eta (t) \eta (t') \rangle = 2D \delta (t - t')
\label{white}
\end{equation}
where here the diffusion constant $D$ is defined as $\alpha k_B T$.

This white noise limit was the original system
investigated by Kramers in his seminal paper \cite{kra}. The system was a
model for chemical reactions, in a which particle overcame
its original bonds to form stronger bonds, by activation caused by random
molecular collisions.

In the last few years there has been a great deal of work done in investigating
a non-Markovian generalisation of (\ref{whlang}) in which the noise is no
longer
white, but the frictional term is still of the form $\alpha \dot{x}$ --- there
is no memory kernel as in (\ref{langevin}). The physical origin of stochastic
equations of this type is very different to the origin of equations of the
type (\ref{langevin}), as described above. In the latter case the noise arose
from the existence of a thermal bath and the strength of the noise was directly
related to the temperature of the bath. The fact that the system plus the bath
are described by the equations of classical mechanics ensure that the same
function which occurs in the dissipative term of (\ref{langevin}) also occurs
in the characterisation of the fluctuations in (\ref{noise}). This is the
fluctuation-dissipation theorem, and noise of this type is called internal,
because it is produced by the interaction of the system with its environment.
The former type of noise, where the Langevin equation has its white noise
form, but where the noise is non-white is called external noise (for reviews
see \cite{ref1} and \cite{ref2}). In this
case the production of the noise is independent of the system, and thus
there will be no fluctuation-dissipation theorem; the noise spectrum is
simply a given function which is independent of the details of the system on
which it acts. The simplest example of external noise acting on a model
system is described by the Langevin equation
\begin{equation}
m\ddot{x}(t) + \alpha \dot{x}(t) + V'(x(t)) = \xi (t)
\label{external}
\end{equation}
where the noise is Gaussian with zero mean and correlator
\begin{equation}
\langle \xi(t)\xi(t')\rangle = \frac{D}{\tau}\exp \left( -\frac{|t-t'|}{\tau}
\right)
\label{excorr}
\end{equation}
Here $D$ is the strength of the noise, not in general related to
temperature. In the limit $\tau \rightarrow 0$ the correlation function
reduces to the white noise delta function shown in (\ref{white}).

Much of the progress in the investigation of the behaviour of systems
which are acted on by external noise has come from the realisation that
physical quantities of interest, such as the stationary probability
distribution (SPD), the escape rate over a potential barrier, correlation
functions and response functions, can be
systematically calculated in the weak noise limit. This is achieved by
applying the method of steepest descent to a path-integral representation
for the conditional probability distribution in the limit $D \rightarrow 0$
\cite{ajm},\cite{tjn}.

The purpose of this paper is to use the same techniques, but on systems
which are subject to internal noise. As we will show, these techniques
can be applied, but unlike the case of external noise, the results are not so
interesting for quantities such as escape rates and the SPD.
They are useful, however, for correlation and response functions and for
conditional probability distributions. The reason for this difference
is clear: whereas for external noise the SPD, and other related quantities
of interest, are unknown a priori, for internal noise the SPD is simply the
Boltzmann distribution. However, the description of the system in terms of
optimal paths in the $D \rightarrow 0$ limit is necessary when trying to find
other quantities, since they cannot simply be found by quadrature.

The outline of the paper is as follows. In section 2 we show how the
Fokker-Planck equation and the path-integral representation of the solution
of this equation, can be formulated for systems with internal noise which
is exponentially correlated. The steepest descent evaluation of the
path-integral is discussed in section 3, where subtleties relating to the
nature of the optimum trajectories in the $m \rightarrow 0$ limit are also
explored. The extension of these ideas to more general non-white noises
is carried out in section 4 and our conclusions are given in section 5.

\section{Formalism}

The most straightforward method of obtaining the Fokker-Planck equation
which is equivalent to the system described by (\ref{external}) and
(\ref{excorr}) is simply to notice that this process can be written as a
Markov process by introducing additional variables. Specifically it can be
expressed as three coupled first order differential equations \cite{ajm}
\begin{equation} \label{eq:ex1}
\dot{x}=u
\end{equation}
\begin{equation} \label{eq:ex2}
m\dot{u}=-\alpha u - V'(x) + \xi
\end{equation}
\begin{equation} \label{eq:ex3}
\tau\dot{\xi}=-\xi+\eta(t)
\end{equation}
where now $\eta(t)$ is Gaussian white noise of strength $D$. This is easy to
verify simply by solving (\ref{eq:ex3}), provided that the initial conditions
are set in the infinitely distant past. In order to make progress with the
analysis of the system described by (\ref{langevin}) and (\ref{noise}), it
is natural to seek an equivalent Markov process, by analogy with the
external noise case. Such a set of equations is easily found. They are
\begin{equation} \label{eq:in1}
\dot{x}=u
\end{equation}
\begin{equation} \label{eq:in2}
m\dot{u}=-V'(x)+\xi
\end{equation}
\begin{equation} \label{eq:in3}
\tau\dot{\xi}=-\alpha u-\xi+\eta(t)
\end{equation}
where again $\eta(t)$ is Gaussian white noise of strength $D$. The only
difference with the external noise set is that the term $-\alpha u$ is
in a different position. This change, although seemingly minor, will turn
out to be very important.

Equations (\ref{eq:in1}-\ref{eq:in3}) fall into the generic class of
Langevin equations which may be written as
\begin{equation}
\dot{z}_i = A_i (\underline{z}) + \eta _i (t)
\label{gen1}
\end{equation}
where $\underline{z}=(x,u,\xi )$ and where
\begin{equation}
\langle \eta_i (t) \eta_j (t') \rangle = 2 B_{ij} \delta (t - t')
\label{gen2}
\end{equation}
Notice, however, that in this case the diffusion matrix $B$ is singular: it
has only one non-zero entry, namely, $B_{33}=D\tau ^{-2}$. The Fokker-Planck
equation for the conditional probability distribution
$P(\underline{z},t|\underline{z}_0,t_0)$ can be found by standard methods
\cite{ris}. In terms of the variables of the problem of interest to us here
it reads
\begin{equation} \label{eq:infp}
\frac{\partial P}{\partial t}=
-\frac{\partial}{\partial x}\left[ uP \right]
+m^{-1}\frac{\partial}{\partial u}\left[\left(V'(x)-\xi\right) P\right]
+\tau^{-1}\frac{\partial}{\partial \xi}\left[\left( \alpha u+\xi\right) P
\right]
+\frac{D}{\tau^{2}}\frac{\partial^{2} P}{\partial \xi^{2}}
\end{equation}
Unlike the external noise problem the stationary probability distribution for
the internal noise problem may be found by separation of variables and is
\begin{equation} \label{eq:inpst}
P_{st}(x,u,\xi)={\cal C}\exp\left( -\frac{\alpha}{D} \left[
\frac{1}{2}mu^{2}+V'(x)+\frac{\tau}{2\alpha}\xi^{2} \right] \right)
\end{equation}
Integrating (\ref{eq:inpst}) over $\xi$ leads to the Maxwell-Boltzmann
distribution as required. However our aim in this section is to write the
solution of (\ref{eq:infp}) in the form of a path-integral, so that we
can use approximate methods to study the problem further. To do this we
can parallel the discussion for the external noise problem \cite{ajm},
given the similarities between (\ref{eq:ex1}-\ref{eq:ex3}) and
(\ref{eq:in1}-\ref{eq:in3}). We begin by noticing that the set of three
first-order stochastic differential equations for internal noise may be
written as one third-order stochastic differential equation:
\begin{equation}
m\ddot{x} + \alpha \dot{x} + V'(x) + \tau (m\xtdots + \dot{x} V''(x) )
= \eta (t)
\label{third}
\end{equation}
Using (\ref{third}) to define a mapping from the noise $\eta$ to the
coordinate $x$, we can transform the probability density functional for
the noise, which is given by
\begin{equation} \label{eq:pdfwhite}
P[\eta]={\cal N}\exp -\int_{t_{0}}^{t}dt\eta ^{2}(t)/4D
\end{equation}
to a probability density functional for the coordinate:
\begin{equation} \label{eq:pdfx}
P[x]={\cal N}J[x]\exp-S[x]/D
\end{equation}
where $S[x]$ is a generalised action functional for internal coloured noise
given by
\begin{equation} \label{eq:s}
S[x]=\frac{1}{4}\int_{t_{0}}^{t}dt
\left[m\ddot{x}+\alpha\dot{x}+V'(x)+\tau(m\xtdots +\dot{x}V''(x))\right]^{2}
\end{equation}
and $J[x]$ is a Jacobian factor given by
\begin{equation} \label{eq:j}
J[x]={\cal C}\exp\left( \frac{1}{2}\int_{t_{0}}^{t}
\tau^{-1}~dt
\right)
\end{equation}
Probability distributions, correlation functions and other quantities of
interest can be found by integration of the appropriate
functions over paths $x(t)$ with weight (\ref{eq:pdfx}).  In the limit of
$D\rightarrow 0$ these path-integrals can be evaluated by the method of
steepest descents; the paths which dominate the integrals being the ones for
which $\delta S[x]/\delta x=0$. This will be the subject of the next section,
but let us first mention other studies on the effect of non-white internal
noise.

Two general approaches are possible: one can investigate the non-Markov
process directly \cite{gah,mor,kub,jac} or use the equivalent Markov process
\cite{im,dms,tal}. The majority of authors consider the diffusion over a
piecewise parabolic potential, but simulations \cite{sbb} have shown the
results for these are unreliable for large damping. In ref. \cite{im} the
method of matrix continued fractions \cite{ris} has been used to solve the
equivalent Fokker-Planck equations and the expressions which are found for the
Fourier transform of the velocity autocorrelation function agree well with the
analogue simulations performed in \cite{ims}.

\section{The weak noise limit}

To evaluate path-integrals with a probability density given by (\ref{eq:pdfx})
in the limit $D \rightarrow 0$, one first has to determine the paths which
dominate the path-integral in this limit. These are found by calculating the
functional derivative of (\ref{eq:s}) with respect to $x(t)$ and setting it
equal to zero in an exactly analogous way to previous work on external noise
\cite{tjn}. Since this equation has no explicit time dependence, one can
proceed in the usual way and multiply by $\dot{x}$ and integrate with respect
to time. If we wish to calculate quantities such as escape rates, the
stationary
probability distribution and correlation functions, the required paths are
those over an infinite time interval, as will be discussed shortly, and
these have to begin at an extremum of the potential with the velocity, and
higher derivatives, equal to zero. This means that the integration constant
is zero, and so the resulting equation for the dominant paths is
\[
2m^{2}\dot{x}
\xtdots
-m^{2}\ddot{x}^{2}-\alpha^{2}\dot{x}^{2}+V'^{2}
+2m\dot{x}^{2}V''=2\alpha\tau \left(
2m\dot{x}\xtdots -m\ddot{x}^{2}+\dot{x}^{2}V'' \right)
\]
\begin{equation} \label{eq:dsdx}
+\tau^{2} \left( m^{2}(2\dot{x}
\xfidots
-2\ddot{x}
\xfodots
+
\xtdots^{2})
+\dot{x}^{2}V''^{2}+m(4\dot{x}
\xtdots
V''-2\ddot{x}^{2}V''
+4\dot{x}^{2}\ddot{x}V'''+2\dot{x}^{4}V'''') \right)
\end{equation}
An exact solution for the above equation can easily be written down by
analogy with the $m,\tau =0$ case \cite{tjn}, even when $m,\tau\neq 0$.
This is simply because there will always be a solution corresponding to
the situation where the optimum value of the noise $\eta $ is zero. The
solution is
\begin{equation} \label{eq:soldown}
m\ddot{x}+\alpha\dot{x}+V'(x)+\tau(m\xtdots +\dot{x}V''(x))=0
\end{equation}
and we expect it to describe transitions where there is a monotonic decrease
in $V(x)$ from the initial state to the final state, since no noise is
required to activate the process. This will therefore be referred to as the
``downhill" solution, and from (\ref{eq:s}) and (\ref{eq:soldown}) it is
clear that it has zero action. However, it is also clear that equation
(\ref{eq:dsdx}) is unchanged by the transformations $\tau\rightarrow -\tau$ and
$\alpha\rightarrow -\alpha$, and so another solution is
\begin{equation} \label{eq:solup}
m\ddot{x}-\alpha\dot{x}+V'(x)-\tau(m\xtdots +\dot{x}V''(x))=0
\end{equation}
This solution certainly does not have zero action, but it is the solution
of least action that satisfies the appropriate boundary conditions. To see
this let us suppose that we have found a solution of (\ref{eq:dsdx}) and
that we write it in the form
\begin{equation}
m\ddot{x}-\alpha\dot{x}+V'(x)-\tau(m\xtdots +\dot{x}V''(x))-R(t)=0
\label{gensol}
\end{equation}
where $R(t)$ is some function of time which characterises the solution.
Substituting (\ref{gensol}) into (\ref{eq:s}) gives the action of this
solution to be, after some algebra,
\begin{equation} \label{eq:sols}
S=\alpha \left[ \frac{1}{2}m\dot{x}^{2}+V(x) \right]_{t_{0}}^{t}
+\frac{1}{2}\tau \left[ (m\ddot{x}+V'(x))^{2} \right]_{t_{0}}^{t}
+ \frac{1}{4}\int^{t}_{t_0}dt R^{2} (t)
\end{equation}
Here it is understood that this is the action on some segment of an infinite
time trajectory. But $R=0$ is a solution (given by (\ref{eq:solup})), so that
if this solution satisfies the appropriate boundary conditions, then
(\ref{eq:sols}) shows that it is the solution of the form (\ref{gensol})
which has the least action. From equations (\ref{eq:in1}) and (\ref{eq:in2})
the action of the $R=0$ solution can be written as
\begin{equation} \label{eq:solsa}
S=\left[  \alpha(\frac{1}{2}mu^{2}+V(x)) +\frac{\tau}{2}\xi^{2}
\right]_{t_{0}}^{t}
\end{equation}
The interpretation of (\ref{eq:solsa}) depends on exactly what quantity is
being calculated. For example, if one wished to find the SPD one would take
$t_0 \rightarrow -\infty$ so that
$S=\alpha (mu^2 /2 + V(x)) + \tau \xi ^2 /2$, up to a constant. This
is exactly the form of the exponent of the SPD obtained in section 2 by
direct integration of the Fokker-Planck equation. On the other hand, if one
wished to calculate the escape rate from a potential well, one
would be interested in paths which took an infinite time to
interpolate between stable and unstable points of the potential
\cite{tjn}. These paths have $V'(x)=0$ and $\ddot{x}=\dot{x}=0$ at their
endpoints, and from (\ref{eq:in1}) and (\ref{eq:in2}) therefore also have
$u=0$ and $\xi =0$. In this case the action (\ref{eq:solsa}) becomes
$\alpha \Delta V$, where $\Delta V$ is the difference between the value of
the potential at the end of the path and that at the beginning i.e. it is
the barrier height. Therefore, to leading order, the average escape rate
is $\exp(-\alpha\Delta V/D)$, which is identical to the white noise result.
This is another consequence of the fluctuation-dissipation theorem.

These examples should not lead one to believe, however, that all quantities
that one might wish to calculate can be found to leading order from very
general properties of the system. Most cannot; it is necessary to solve
the differential equation for $x(t)$ in order to find the leading order
behaviour of most quantities. In fact, in some cases it may be necessary
to solve the original sixth-order equation $\delta S[x]/\delta x =0$,
rather than (\ref{eq:dsdx}), since the initial and final conditions on the
path may not imply that the constant of integration required to derive
(\ref{eq:dsdx}) is zero. An example is in the calculation of the
conditional probability density itself. In general the values of $x$ and
its derivatives that are specified at the initial and final times are not
compatible with a path of zero ``energy". Another example where the
detailed solution of the path is required is in the calculation of
correlation functions such as $\langle x(t_1 ) x(t_2 )\rangle$. In this case
the zero trajectories are the relevant ones and so (\ref{eq:dsdx}) may be
used. Since such a calculation would necessitate developing the formalism to
the next order, we will not pursue this any further here.

So far we have shown that the dominant, or optimal, paths are given by the
solution of the differential equation (\ref{eq:dsdx}). We will now show that
care has to be taken in the limit $m\rightarrow 0$; unlike the case of
external noise the paths are singular in this limit. Let us first of all
naively set $m=0$ in this equation. Then
\begin{equation}\label{eq:pathm0}
\alpha \dot{x}=\pm\frac{V'(x)}{1+\frac{\tau}{\alpha}V''(x)}
\end{equation}
This expression shows that there is the possibility of the velocity becoming
infinite. We can see this more clearly by taking $V(x)$ to have a definite
form, such as the double well potential $V(x)=-x^{2}/2 + x^{4}/4$. In this
case if $\tau /\alpha < 1$, the denominator of (\ref{eq:pathm0}) is
always positive, and hence if we are considering a path where
$\dot{x} > 0$, then we choose the positive (negative) sign in
(\ref{eq:pathm0}) if $V'(x)$ is positive (negative), and vis-versa if
$\dot{x} < 0$. This is the familiar situation found in the  white-noise
dominant paths. But now there is the possibility that if
$\tau /\alpha > 1$, the denominator diverges for certain values of
$x$. This would also mean that, since the denominator would be negative
for some values of $x$, the minus sign would have to be taken in
(\ref{eq:pathm0}), even if both $\dot{x}$ and $V'(x)$ were positive. For
%
%\begin{figure}[h]
%\begin{center}
%\mbox{
%\epsfysize=10cm
%\epsfxsize=\textwidth
%\epsfbox{fig1.eps}
%}
%\end{center}
%\caption{Sketch of double well potential}
%\end{figure}
%
example consider the path from $x=-1$ to $x=+1$ where the velocity $\dot{x}$
is positive. Figure 1 shows the double-well potential with stable minima at
$a$ ($x=-1$) and $e$ ($x=+1$), and a maximum at $c$ ($x=0$). The
points $b$ and $d$ are where
the denominator in (\ref{eq:pathm0}) changes sign, that is, where the velocity
becomes infinite. Therefore for the regions $a\rightarrow b$ and
$c\rightarrow d$ we have to take the positive sign in (\ref{eq:pathm0}) and
the negative sign for the other two regions. Since it seems physically
unreasonable to have an ``uphill" path giving zero action and a ``downhill"
path giving a non-zero action, we conclude that it is incorrect to set $m=0$
in the differential equations for the solutions.

This can be seen in another way. Suppose we set $m=0$ in the equivalent
Markov process (\ref{eq:in1})-(\ref{eq:in3}). After rescaling time so that
$t\rightarrow\alpha t$, these three equations can be combined to give:
\begin{equation}\label{eq:l1}
\left(1+\frac{\tau}{\alpha}V''(x) \right)\dot{x}+V'(x)=\eta
\end{equation}
Making the substitution
\begin{equation}\label{eq:y}
y=x+\frac{\tau}{\alpha}V'(x)
\end{equation}
gives the equation
\begin{equation}\label{eq:yl}
\dot{y}+\frac{\partial U(y)}{\partial y}=\eta
\end{equation}
where
\begin{equation}\label{eq:uy}
U(y) \equiv V(x(y))+\frac{\tau}{2\alpha}V'^{2}(x(y))
\end{equation}
and where $x(y)$ is found by inverting (\ref{eq:y}). The Langevin equation
(\ref{eq:l1}) has now been transformed into a simpler problem: an overdamped
particle moving in a potential $U(y)$ and subject to white noise. A
parametric plot of $U(y)$ against $y$ for differing values of $\tau /\alpha$
is shown in figure 2.
%
%\begin{figure}[h]
%\begin{center}
%\mbox{
%\epsfysize=10cm
%\epsfxsize=\textwidth
%\epsfbox{fig2.eps}
%}
%\end{center}
%\caption{Potential U(y) against y}
%\end{figure}
%
For $\tau /\alpha < 1$ the potential is single valued and a calculation of
the action gives a value of 1/4. However if $\tau /\alpha >1$, the potential
is double valued and the above transformation is ill-defined.

We have given two illustrations of the fact that setting $m$ equal to zero
in the equations for the dominant paths is a singular procedure. We will
now, by investigating the nature of the solutions to the ``uphill" path
(\ref{eq:solup}) for small $m$, show that the paths are well-defined
for any finite $m$, and it is only in the limit $m=0$ that problems occur.
We will do this by exploiting the fact that the desired solution of
(\ref{eq:solup}) grows near the minimum $a$ and decays near the maximum $c$.
Since this only requires analysis of the equation near extrema of the
potential, it is sufficient to investigate the linearised form of the
equation, that is, to assume a parabolic form for $V(x)$:
\begin{equation}\label{eq:vac}
V(x)\approx V(x_{0}) + \frac{1}{2}V''(x_{0})(x-x_{0})^{2}
\end{equation}
where ``$0$" denotes either $a$ or $c$. From the linearised equation the
solution is easily seen to have the form
$x_{0}+ C_0 e^{\lambda_{0} t}$, where $C_0$ is a constant and $\lambda _0$
satisfies the equation
\begin{equation}\label{eq:le}
(m\lambda_{0}^{2}+V''(x_{0}))(\tau\lambda_{0}-1)=-\alpha\lambda_{0}
\end{equation}
or using the explicit values $V''(a)=2$ and $V''(c)=-1$
\begin{eqnarray}
m\tau\lambda ^3_a - m\lambda ^2_a +
(\alpha + 2\tau )\lambda _a - 2 & = & 0 \\
m\tau\lambda ^3_c - m\lambda ^2_c + (\alpha - \tau )\lambda _c + 1 & = & 0
\label{root}
\end{eqnarray}
To show that an uphill path exists it is necessary to show that there exist
solutions of these equations which have $\Re \lambda _a > 0$ and
$\Re \lambda _c < 0$. In the limit $m \rightarrow 0$, the cubic equation for
$\lambda _a$ has one real and two complex conjugate roots with the real parts
of these roots having a finite, positive limit at $m=0$. The equation for
$\lambda _c$ has only one root with a negative real part for small $m$. But
while this has a finite $m=0$ limit if $\tau /\alpha < 1$, it does not if
$\tau /\alpha \ge 1$. Specifically
\begin{equation}
\lambda _c = \left\{ \begin{array}{ll}
-\frac{1}{(m\tau )^{1/3}} + \frac{1}{3\tau } + O(m^{1/3})
& \mbox{if $\tau /\alpha = 1$} \\
- \frac{1}{\sqrt{m}}\left[ \frac{\tau - \alpha }{\tau}
\right]^{1/2} - \frac{\alpha}{2\tau (\tau - \alpha )} + O(m^{1/2})
& \mbox{if $\tau /\alpha > 1$}
\end{array}
\right.
\label{lamroot}
\end{equation}
Therefore in the limit $m=0$ there is no solution of the linearised form
of (\ref{eq:solup}) which is exponentially decaying near the point $c$.
The linearised form of (\ref{eq:soldown}) may be analysed in a similar way.
Since it is just (\ref{eq:solup}) with the substitution
$\tau \rightarrow -\tau$ and $\alpha \rightarrow -\alpha$, the cubic
equations for $\lambda _0$ are unchanged if we let
$\lambda _0 \rightarrow -\lambda _0$. Hence the above analysis is
unchanged apart from replacing $\lambda _c$ by $-\lambda _c$ and $\lambda _a$
by $-\lambda _e$. Once again solutions exist which have the correct
behaviour as $m \rightarrow 0$ near the bottom of the potential wells, but
there is no exponentially growing solution near the point $c$ when $m=0$.

\section{Generalisations}

In the earlier sections of this paper we have been investigating systems
described by an equation of the form (\ref{langevin}) with the memory
kernel given by $f(t)=\alpha \tau ^{-1} \exp-|t|/\tau$. This is because,
as we shall see below, this is the simplest non-trivial choice for $f$.
In this section we wish to show how the formalism we have discussed can
be extended to more general memory functions.

We begin by writing (\ref{langevin}) as
\begin{equation} \label{eq:genin3}
\dot{x}=u
\end{equation}
\begin{equation} \label{eq:genin4}
m\dot{u}=-\int_{-\infty}^{\infty}dt'g(t-t') u(t') - V'(x) + \xi (t)
\end{equation}
where
\begin{equation} \label{eq:geninr}
g(t-t') =\left\{
\begin{array}{ll}
f(t-t') & t>t'    \\
0       & t<t'
\end{array}
\right.
\end{equation}
This condition can be more simply expressed by the use of Fourier transforms:
\begin{equation} \label{eq:genincorw}
\tilde{f}(\omega)=\int_{-\infty}^{\infty}dte^{i\omega t}f(|t|)
=\int_{0}^{\infty}dte^{i\omega t}f(t)~+~c.c
\end{equation}
where $c.c.$ is the complex conjugate, and
\begin{equation} \label{eq:geninrw}
\tilde{g}(\omega)=\int_{-\infty}^{\infty} g(t)
e^{i\omega t}=\int_{0}^{\infty}dtf(t)e^{i\omega t}
\end{equation}
{}From (\ref{eq:genincorw}) and (\ref{eq:geninrw}) we have that
\begin{equation} \label{eq:geninrcw}
\tilde{f}(\omega)=\tilde{g}(\omega)+c.c.
\end{equation}
Returning now to the set of equations (\ref{eq:genin3}) and (\ref{eq:genin4}),
we rewrite them trivially as
\begin{equation} \label{eq:g1}
\dot{x}=u
\end{equation}
\begin{equation} \label{eq:g2}
m\dot{u}=-V'(x)+\zeta
\end{equation}
\begin{equation} \label{eq:g3}
\zeta =-\int_{-\infty}^{\infty}dt' g(t-t') u(t') + \xi (t)
\end{equation}
The advantage of this method of expressing the problem is that the last
equation, (\ref{eq:g3}), is linear in the variables $\zeta ,u$ and $\xi$
and so may be Fourier transformed to yield the simpler expression:
\begin{equation} \label{eq:g3w}
\tilde{\zeta}(\omega)=-\tilde{u}(\omega)
\tilde{g}(\omega) + \tilde{\xi}(\omega)
\end{equation}
where the correlation function of the noise $\tilde{\xi}(\omega)$ in frequency
space is given by the Fourier transform of equation (\ref{noise}):
\begin{equation} \label{eq:fcorw}
\langle \tilde{\xi}(\omega)\tilde{\xi}(\omega') \rangle =
D\alpha ^{-1} \tilde{f}(\omega)2\pi\delta(\omega+\omega')
\end{equation}
As a check on the above, one can show that it is easy to recover our previous
results for exponentially correlated noise. In this case,
\begin{equation}
\tilde{g}(\omega)=\frac{\alpha}{1-i\omega \tau}
\label{ftg}
\end{equation}
and
\begin{equation}
\tilde{f}(\omega)=\frac{2\alpha}{1+\omega ^{2} \tau ^{2}}
=\frac{2}{\alpha} \tilde{g}(\omega) \tilde{g}(-\omega)
\label{ftf}
\end{equation}
{}From (\ref{eq:fcorw}) and (\ref{ftf}) we see that if we define a new
noise by $\tilde{\eta}(\omega)=[1-i\omega\tau ] \tilde{\xi}(\omega)$
then $\eta$ is a white noise with correlation function given by
(\ref{white}), and (\ref{eq:g3w}) and (\ref{ftg}) now give
\begin{equation}
[1-i\omega \tau ]\tilde{\zeta}(\omega) =
-\alpha \tilde{u}(\omega) + \tilde{\eta}(\omega)
\label{zeta}
\end{equation}
Taking the inverse Fourier transform of (\ref{zeta}) gives (\ref{eq:in3}),
as required.

The starting point in setting up the path-integral formalism from any
Langevin equation, is the probability density functional for the noise. In
the most general case under consideration here it is Gaussian with zero
mean and correlator given by (\ref{noise}) with $k_B T=D/\alpha$, hence
this functional is
\begin{equation}
P[\xi ] \propto \exp -\alpha \int d\omega \tilde{\xi}^{*} (\omega)
\tilde{f}^{-1}(\omega) \tilde{\xi}(\omega) /2D
\label{path}
\end{equation}
This is, of course, exactly the same starting point that is used to study
problems with arbitrary external noise, and many of the comments we will
now make have analogues in that case \cite{ajm2}. We first note that if we
are to define the noise through (\ref{path}), then $\tilde{f}^{-1}(\omega)$
has to be positive for all $\omega$ to ensure convergence of path
integrals. Thus $\tilde{f}^{-1}(\omega)$ must have no real zeros. We now
ask that $\tilde{f}^{-1}(\omega)$ can be written as a power series in
$\omega ^2$ (since $f(t)$ is even in $t$, $\tilde{f}(\omega)$ is even in
$\omega$). This is so that it is straightforward to write the exponent in
(\ref{path}) in terms of $\xi (t)$ and its derivatives, rather than
$\tilde{\xi}(\omega )$. This then enables us to eliminate $\xi (t)$and its
derivatives in favour of $x(t)$ and its derivatives using the Langevin
equation.

It is now clear that the simplest processes are those for which the power
series $\tilde{f}^{-1}(\omega)$ terminates. If it terminates after the term
of order $\omega ^{2n}$ we shall call it an $n$th order process \cite{ajm2}.
Since this is a polynomial which is even in $\omega$ and has no real zeros,
we may write
\begin{equation} \label{eq:genincorwma}
\tilde{f}^{-1}(\omega)=\frac{1}{2\alpha}\prod_{m=1}^{n}
\left[ 1+\alpha_{m}^{2}\omega^{2} \right]
\end{equation}
or
\begin{equation} \label{eq:geninc}
\tilde{f}(\omega)=\frac{2\alpha}{\prod_{m=1}^{n} \left[ 1-i\alpha_{m}\omega
\right] \left[ 1+i\alpha_{m}\omega \right]}
\end{equation}
where $Re(\alpha_{m})\neq 0$; $m=1,2,\ldots,n$. The simplest possible process
is white noise, with $\tilde{f}(\omega)=2\alpha$. The simplest non-Markov
process is the $n=1$ process which, by comparing (\ref{ftf}) and
(\ref{eq:geninc}), we see is a process with exponentially correlated noise
where $\alpha_1 =\tau$.

For an $n=2$ process, if we define
\begin{equation}
\tilde{\eta}(\omega)=
[1-i\alpha _1\omega][1-i\alpha _2\omega]\tilde{\xi}(\omega)
\label{eta}
\end{equation}
then $\eta$ is a white noise. Equations (\ref{eq:geninc}) and
(\ref{eq:geninrcw}) now give for (\ref{eq:g3w})
\begin{equation}
[1-i\alpha _1\omega][1-i\alpha _2\omega]\tilde{\zeta}(\omega)
=-\tilde{u}(\omega)\alpha [1-i\beta _2\omega] + \tilde{\eta}(\omega)
\label{ntwo}
\end{equation}
where
\begin{equation}
\beta _2 = \frac{\alpha _1 \alpha _2}{\alpha _1 + \alpha _2}
\label{beta2}
\end{equation}
Writing (\ref{ntwo}) as
\begin{equation}
-i\omega \beta _2 (\alpha \tilde{u}(\omega)
-i\omega \beta _1 \tilde{\zeta}(\omega)) = -\tilde{\zeta}(\omega) -
(\alpha \tilde{u}(\omega) -i\omega \beta _1 \tilde{\zeta}(\omega))
+ \tilde{\eta}(\omega)
\label{ntwox}
\end{equation}
where
\begin{equation}
\beta _1 = \alpha _1 + \alpha _2
\label{beta1}
\end{equation}
we see that by defining
\begin{equation}
\tilde{\rho}(\omega) = \alpha \tilde{u}(\omega) -i\omega \beta _1
\tilde{\zeta}(\omega)
\label{rho}
\end{equation}
we may rewrite (\ref{eq:g1}-\ref{eq:g3}) for the general $n=2$ process as
a set of four coupled first-order equations:
\begin{equation}
\dot{x} = u
\label{f1}
\end{equation}
\begin{equation}
m\dot{u} = -V'(x) + \zeta
\label{f2}
\end{equation}
\begin{equation}
\beta _1\dot{\zeta} = -\alpha u + \rho
\label{f3}
\end{equation}
\begin{equation}
\beta _2 \dot{\rho} = -\zeta -\rho +\eta (t)
\label{f4}
\end{equation}
We can now use this to construct the equivalent Fokker-Planck equation and its
path-integral solution, in an exactly analogous way to the procedure we used
when starting from (\ref{eq:in1}-\ref{eq:in3}). That is, we note that
(\ref{f1}-\ref{f4}) may be written in the form (\ref{gen1}), with noise
correlators given by (\ref{gen2}). Hence the Fokker-Planck equation for
the conditional density distribution is
\begin{equation} \label{eq:m2fp}
\frac{\partial P}{\partial t}=
-\frac{\partial [uP]}{\partial x}
+m^{-1}\frac{\partial \left[(V'(x)-\zeta)P\right]}{\partial u}
+\beta_{1}^{-1}\frac{\partial \left[(\alpha u-\rho)P\right]
}{\partial \zeta}
+\beta_{2}^{-1}\frac{\partial \left[(\zeta+\rho)P\right]
}{\partial \rho}
+\frac{D}{\beta_{2}^{2}}\frac{\partial^{2}P }{\partial \rho ^{2}}
\end{equation}
The stationary solution has the separable form
\begin{equation} \label{eq:m2fpst}
P_{st}(x,u,\zeta,\rho)={\cal N}\exp \left\{
-\frac{\alpha}{D} \left[ \frac{1}{2}mu^{2}+V(x)+\frac{\beta_{1}}{2\alpha}
\zeta ^{2}+\frac{\beta_{2}}{2\alpha}\rho ^{2} \right] \right\}
\end{equation}
To write the solution of (\ref{eq:m2fp}) as a path-integral, we first note
that the a set of four first-order equations (\ref{f1}-\ref{f4}) may be
written as one fourth-order equation:
\begin{equation}
m\ddot{x}+\alpha\dot{x}+V'
+(\beta_{1}m\xtdots+\beta_{1}\dot{x}V''+\alpha\beta_{2}\ddot{x})
+\beta_{1}\beta_{2}(m\xfodots+\ddot{x}V''+\dot{x}^{2}V''') = \eta (t)
\label{fourth}
\end{equation}
This is exactly analogous to (\ref{third}). From now on the construction
of the path-integral follows the discussion in section 2.

It is clear how to extend the above to a general $n$th-order process. One
finds that the equivalent Markov process is a simple generalisation of
(\ref{eq:in1}-\ref{eq:in3}) and (\ref{f1}-\ref{f4}), but with $(n+2)$
first-order equations. Only the second one, containing the potential term,
is non-linear, and only the last one contains the white noise term. It is
therefore possible to write this set as a single $(n+2)$th-order stochastic
differential equation with white noise. From this point on the construction
of the path-integral is again standard.

If $\tilde{f}^{-1}(\omega)$ is not a polynomial in $\omega ^{2}$, it may
still be possible to approximate it by one. For instance, if the noise is
characterised by a single time scale, $\tau$, then the coefficients
$\{\alpha _m\}$ are of order $\tau$, and hence the generalisation of the
$\beta _1$ and $\beta _2$ coefficients that appear for the $n=2$ case above
are also of order $\tau$. For small $\tau$ one can then argue that the higher
order terms in the generalisations of (\ref{third}) and ({\ref{fourth}) can
be neglected, that is, the full function $\tilde{f}^{-1}(\omega)$ may be
approximated by a polynomial of an appropriate order. This fact has been
used to calculate the mean first passage time as a perturbative expansion
for small $\tau$ in an infinite order external noise process \cite{ajm2}.

In this section we have concentrated on explaining how the methods used in
the earlier part of the paper can be extended to a system with one
coordinate $x$, but a very general form for the noise. The generalisation
to the case where the system itself has many degrees of freedom is
straightforward; the Langevin equation which replaces (\ref{langevin}),
and which we use as our starting point, has a very similar structure
\cite{zwa} and most of the formalism we have discussed in this paper goes
through essentially unchanged. Finally. as shown in \cite{zwa} the noise
will, in general, be multiplicative. The whole formalism can be adapted to
this situation too, just as it can for external noise \cite{tjn}.

\section{Conclusions}

In this paper we have determined the paths that dominate in the weak-noise
limit of stochastic processes where the noise is internal and coloured. For
paths that have a finite action over an infinite time interval, the value
of the action is always $\alpha \Delta V$, that is, it is independent of
details of the potential, and only dependent on the height of the barrier
and the constant $\alpha$. Thus the leading order behaviour of quantities
such as the stationary probability distribution is immediately seen to be
given by the Boltzmann distribution, as expected. This form for the action
also means that more exotic phenomena, such as thermal ratchets \cite{mag},
will not exist for systems where the noise is internal. This follows from
the fact that these rachets rely on the escape rates to go from right to
left and left to right over an asymmetric potential barrier differ, because
the actions differ if the noise is coloured and external. In our case the
fluctuation-dissipation theorem, which is a reflection of the feedback
between the system and the environment, ensures that the action is the
same going either way, even if the potential is asymmetric, and hence
the effect cannot occur. Although some physical quantities have the
Boltzmann factor at leading order, many do not, and even those that do will
frequently have a non-trivial prefactor which can only be calculated if
the detailed form of the path is known. We also showed that the $m=0$
limit of a process with exponentially correlated internal noise is singular.
{}From the discussion in Section 4 it is also clear that this behaviour also
occurs for more general internal non-white processes. This should not
be too worrying since these systems were initially derived from a
deterministic Hamiltonian which does not exist for massless particles.
This is in contrast to external noise, where there is no reason why
$m$ should not be formally taken to zero. Since for white noise, internal
and external noises are identical, it follows that the $m=0$ limit of
internal white noise is well-defined; only the introduction of colour
generates the singularity.

\section*{Acknowledgements}

SJBE would like to thank the EPSRC for a research studentship. This work
was also funded in part under EPSRC grant GR/H40150.

%\newpage
\section*{Figure Captions}
\begin{enumerate}
\item{Sketch of the bistable potential showing required path}
\item{Parametric plot of the potential $U(y)$ against $y$ for several values of
$\tau$}
\end{enumerate}

%\newpage

%
\newpage
\begin{figure}
\begin{center}
\mbox{
\epsfysize=15cm
\epsfxsize=\textwidth
\epsfbox{intfig1.eps}
}
\\Figure 1
\end{center}
\end{figure}
\begin{figure}
\begin{center}
\mbox{
\epsfysize=15cm
\epsfxsize=\textwidth
\epsfbox{intfig2.eps}
}
\\Figure 2
\end{center}
\end{figure}
\end{document}